\documentclass[letterpaper,twocolumn,10pt]{article}
\usepackage{usenix2019_v3,epsfig,endnotes}

\usepackage{booktabs} 
\usepackage{outlines}
\usepackage{comment}
\usepackage[ruled,vlined]{algorithm2e}

\usepackage{epsfig}
\usepackage{graphics}
\usepackage{subcaption}
\usepackage{caption}
\usepackage{color}
\usepackage{times}
\usepackage{xspace}
\usepackage{listings}
\usepackage{hyperref}
\usepackage{enumitem}
\usepackage{siunitx}
\usepackage{multirow}
\usepackage{tabularx}
\usepackage{booktabs}
\usepackage[available,functional,reproduced]{usenixbadges}

\usepackage[utf8]{inputenc}

\usepackage{authblk}

\DeclareFixedFont{\ttb}{T1}{txtt}{bx}{n}{12} 
\DeclareFixedFont{\ttm}{T1}{txtt}{m}{n}{12}  

\usepackage{color}
\definecolor{deepblue}{rgb}{0,0,0.5}
\definecolor{deepred}{rgb}{0.6,0,0}
\definecolor{deepgreen}{rgb}{0,0.5,0}

\usepackage{listings}

\definecolor{bluekeywords}{rgb}{0.13,0.13,1}
\definecolor{greencomments}{rgb}{0,0.5,0}
\definecolor{turqusnumbers}{rgb}{0.17,0.57,0.69}
\definecolor{redstrings}{rgb}{0.5,0,0}
\lstdefinelanguage{python}
    {
    keywordsprefix=\$,
    numbers=none,
    keywordstyle=\color{turqusnumbers},
    morekeywords=[1]{helper_funct, r0, r1,r2,r3,r4,r5, exit, return, struct, __u32},
    keywordstyle=[1]\bfseries,
    morekeywords=[2]{FPAEvolution, switch_features_handler, function, __init__, add_flow, configure, send_msg, OFPMatch, OFPFlowMod},
    keywordstyle=[2]\color{redstrings},
    morekeywords=[3]{},
    keywordstyle=[3]\bfseries,
    sensitive=false,
    morecomment=[l][\color{greencomments}]{\#},
    morestring=[b]',
    stringstyle=\color{redstrings}
    }
\newcommand{\pproc}[0]{\texttt{Sephirot}\xspace}
\newcommand{\nxdp}[0]{{hXDP}\xspace}

\begin{document}

\date{}

\title{hXDP: Efficient Software Packet Processing on FPGA NICs}

\makeatletter
\renewcommand\AB@affilsepx{, \protect\Affilfont}
\makeatother

\author[1,3]{Marco Spaziani Brunella}
\author[1,3]{Giacomo Belocchi}
\author[1,2]{Marco Bonola}
\author[1]{\rm Salvatore Pontarelli}
\author[4]{Giuseppe Siracusano}
\author[3]{Giuseppe Bianchi}
\author[2,3]{Aniello Cammarano}
\author[2,3]{Alessandro Palumbo}
\author[2,3]{Luca Petrucci}
\author[4]{Roberto Bifulco}

\affil[1]{Axbryd}
\affil[2]{CNIT}
\affil[3]{University of Rome Tor Vergata}
\makeatletter
\renewcommand\AB@affilsepx{,\\ \protect\Affilfont}
\makeatother
\affil[4]{NEC Laboratories Europe}

\renewcommand\Authands{ and }

\maketitle
\pagestyle{empty}

\begin{abstract}
FPGA accelerators on the NIC enable the offloading of expensive packet processing tasks from the CPU. However, FPGAs have limited resources that may need to be shared among diverse applications, and programming them is difficult.

We present a solution to run Linux's eXpress Data Path programs written in eBPF on FPGAs, using only a fraction of the available hardware resources while matching the performance of high-end CPUs.
The iterative execution model of eBPF is not a good fit for FPGA accelerators. Nonetheless, we show that many of the instructions of an eBPF program can be compressed, parallelized or completely removed, when targeting a purpose-built FPGA executor, thereby significantly improving performance. We leverage that to design \nxdp, which includes (i) an optimizing-compiler that parallelizes and translates eBPF bytecode to an extended eBPF Instruction-set Architecture defined by us; a (ii) soft-CPU to execute such instructions on FPGA; and (iii) an FPGA-based infrastructure to provide XDP's maps and helper functions as defined within the Linux kernel.

We implement \nxdp on an FPGA NIC and evaluate it running real-world unmodified eBPF programs. Our implementation is clocked at 156.25MHz, uses about 15\% of the FPGA resources, and can run dynamically loaded programs. Despite these modest requirements, it achieves the packet processing throughput of a high-end CPU core and provides a 10x lower packet forwarding latency.
\end{abstract}

\lstdefinestyle{example}{
  float=tp,
  floatplacement=tbp,
  abovecaptionskip=-5pt
}
\captionsetup[lstlisting]{singlelinecheck=false,margin=3pt,
     font={bf,sf,footnotesize},skip=0pt}

\section{Introduction}
\label{sec:intro}

FPGA-based NICs have recently emerged as a valid option to offload CPUs from packet processing tasks, due to their good performance and re-programmability. Compared to other NIC-based accelerators, such as network processing ASICs~\cite{rmt} or many-core System-on-Chip SmartNICs~\cite{netronome}, FPGA NICs provide the additional benefit of supporting diverse accelerators for a wider set of applications~\cite{ovtcharov2015toward}, thanks to their embedded hardware re-programmability. Notably, Microsoft has been advocating for the introduction of FPGA NICs, because of their ability to use the FPGAs also for tasks such as machine learning~\cite{brainwave, catapult}. FPGA NICs play another important role in 5G telecommunication networks, where they are used for the acceleration of radio access network functions~\cite{intelFPGA5G, xilinxFPGA5G, 9076114, necFPGA5g}. In these deployments, the FPGAs could host multiple functions to provide higher levels of infrastructure consolidation, since physical space availability may be limited. For instance, this is the case in smart cities~\cite{villanueva2013civitas}, 5G local deployments, e.g., in factories~\cite{ricart2018towards,pinneterre2018vfpgamanager}, and for edge computing in general~\cite{jiang2018accelerating,biookaghazadeh2018fpgas}.
Nonetheless, programming FPGAs is difficult, often requiring the establishment of a dedicated team composed of hardware specialists~\cite{AccelNet}, which interacts with software and operating system developers to integrate the offloading solution with the system. Furthermore, previous work that simplifies network functions programming on FPGAs assumes that a large share of the FPGA is dedicated to packet processing~\cite{p4netfpga, p4fpga, FlowBlaze}, reducing the ability to share the FPGA with other accelerators. 

In this paper, our goal is to provide a more general and easy-to-use solution to program packet processing on FPGA NICs, using little FPGA resources, while seamlessly integrating with existing operating systems. We build towards this goal by presenting \texttt{\nxdp}, a set of technologies that enables the efficient execution of the Linux's \texttt{eXpress Data Path} (XDP)~\cite{xdp} on FPGA. XDP leverages the eBPF technology to provide secure programmable packet processing within the Linux kernel, and it is widely used by the Linux's community in productive environments. \nxdp provides full XDP support, allowing users to dynamically load and run their unmodified XDP programs on the FPGA.

The eBPF technology is originally designed for sequential execution on a high-performance RISC-like register machine, which makes it challenging to run XDP programs effectively on FPGA. That is, eBPF is designed for server CPUs with high clock frequency and the ability to execute many of the sequential eBPF instructions per second. 
Instead, FPGAs favor a widely parallel execution model with clock frequencies that are 5-10x lower than those of high-end CPUs. As such, a straightforward implementation of the eBPF iterative execution model on FPGA is likely to provide low packet forwarding performance. 
Furthermore, the \nxdp design should implement arbitrary XDP programs while using little hardware resources, in order to keep FPGA's resources free for other accelerators. 

We address the challenge performing a detailed analysis of the eBPF Instruction Set Architecture (ISA) and of the existing XDP programs, to reveal and take advantage of opportunities for optimization.  
First, we identify eBPF instructions that can be safely removed, when not running in the Linux kernel context. For instance, we remove data boundary checks and variable zero-ing instructions by providing targeted hardware support.  Second, we define extensions to the eBPF ISA to introduce 3-operand instructions, new 6B load/store instructions and a new parametrized program exit instruction. Finally, we leverage eBPF instruction-level parallelism, performing a static analysis of the programs at compile time, which allows us to execute several eBPF instructions in parallel. We design \nxdp to implement these optimizations, and to take full advantage of the on-NIC execution environment, e.g., avoiding unnecessary PCIe transfers. Our design includes: (i) a compiler to translate XDP programs' bytecode to the extended \nxdp ISA; (ii) a self-contained FPGA IP Core module that implements the extended ISA alongside several other low-level optimizations; (iii) and the toolchain required to dynamically load and interact with XDP programs running on the FPGA NIC. 

To evaluate \nxdp we provide an open source implementation for the NetFPGA~\cite{netsume}. We test our implementation using the XDP example programs provided by the Linux source code, and using two real-world applications: a simple stateful firewall; and Facebook's Katran load balancer. \nxdp can match the packet forwarding throughput of a multi-GHz server CPU core, while providing a much lower forwarding latency. This is achieved despite the low clock frequency of our prototype (156MHz) and using less than 15\% of the FPGA resources. 
In summary, we contribute:
\begin{itemize}[noitemsep,topsep=0pt]
    \item the design of \nxdp including: the hardware design; the companion compiler; and the software toolchain;
    \item the implementation of a \nxdp IP core for the NetFPGA
    \item a comprehensive evaluation of \nxdp when running real-world use cases, comparing it with an x86 Linux server. 
    \item a microbenchmark-based comparison of the \nxdp implementation with a Netronome NFP4000 SmartNIC, which provides partial eBPF offloading support.
\end{itemize}
\section{Concept and overview}
\label{sec:overview}
In this section we discuss \nxdp goals, scope and requirements, we provide background information about XDP, and finally we present an overview of the \nxdp design.

\begin{figure}[t]
	\begin{center}
	\includegraphics[width=0.9\columnwidth]{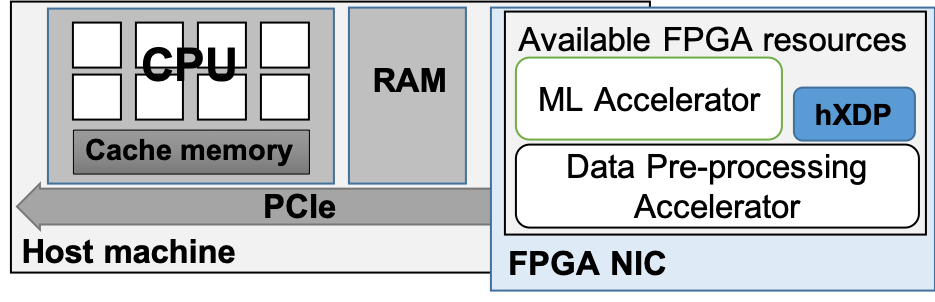} 
	\vspace{-6pt}
	\caption{The \nxdp concept. \nxdp provides an easy-to-use network accelerator that shares the FPGA NIC resources with other application-specific accelerators.} 
	\label{fig:concept}
    \end{center}
	\vspace{-20pt}
\end{figure}

\subsection{Goals and Requirements}
\noindent\textbf{Goals} 
Our main goal is to provide the ability to run XDP programs efficiently on FPGA NICs, while using little FPGA's hardware resources (See Figure~\ref{fig:concept}).

A little use of the FPGA resources is especially important, since it enables extra consolidation by packing different application-specific accelerators on the same FPGA.

The choice of supporting XDP is instead motivated by a twofold benefit brought by the technology: it readily enables NIC offloading for already deployed XDP programs; it provides an on-NIC programming model that is already familiar to a large community of Linux programmers.
Enabling such a wider access to the technology is important since many of the mentioned edge deployments are not necessarily handled by hyperscale companies. Thus, the companies developing and deploying applications may not have resources to invest in highly specialized and diverse professional teams of developers, while still needing some level of customization to achieve challenging service quality and performance levels.
In this sense, \nxdp provides a familiar programming model that does not require developers to learn new programming paradigms, such as those introduced by devices that support P4~\cite{p4} or FlowBlaze~\cite{FlowBlaze}.

\vspace{0.1cm}
\noindent\textbf{Non-Goals} 
Unlike previous work targeting FPGA NICs~\cite{p4fpga, p4netfpga, FlowBlaze}, \nxdp does not assume the FPGA to be dedicated to network processing tasks. Because of that, \nxdp adopts an iterative processing model, which is in stark contrast with the pipelined processing model supported by previous work. The iterative model requires a fixed amount of resources, no matter the complexity of the program being implemented. Instead, in the pipeline model the resource requirement is dependent on the implemented program complexity, since programs are effectively "unrolled" in the FPGA. In fact, \nxdp provides dynamic runtime loading of XDP programs, whereas solutions like P4->NetFPGA~\cite{p4fpga} or FlowBlaze need to often load a new FPGA bitstream when changing application. As such, \nxdp is not designed to be faster at processing packets than those designs. Instead, \nxdp aims at freeing precious CPU resources, which can then be dedicated to workloads that cannot run elsewhere, while providing similar or better performance than the CPU. 

Likewise, \nxdp cannot be directly compared to SmartNICs dedicated to network processing. Such NICs' resources are largely, often exclusively, devoted to network packet processing. Instead, \nxdp leverages only a fraction of an FPGA resources to add packet processing with good performance, alongside other application-specific accelerators, which share the same chip's resources.

Finally, \nxdp is not providing a transparent offloading solution.\footnote{Here, previous complementary work may be applied to help the automatic offloading of network processing tasks~\cite{phothilimthana2018floem}.} While the programming model and the support for XDP are unchanged compared to the Linux implementation, programmers should be aware of which device runs their XDP programs. This is akin to programming for NUMA systems, in which accessing given memory areas may incur additional overheads.  

\vspace{0.1cm}
\noindent\textbf{Requirements} 
Given the above discussion, we can derive three high-level requirements for \nxdp:
\begin{enumerate}[noitemsep,topsep=0pt]
    \item it should execute unmodified compiled XDP programs, and support the XDP frameworks' toolchain, e.g., dynamic program loading and userspace access to maps;
    \item it should provide packet processing performance at least comparable to that of a high-end CPU core;
    \item it should require a small amount of the FPGA's hardware resources. 
\end{enumerate}
Before presenting a more detailed description of the \nxdp concept, we now give a brief background about XDP.

\subsection{XDP Primer}
XDP allows programmers to inject programs at the NIC driver level, so that such programs are executed before a network packet is passed to the Linux's network stack. This provides an opportunity to perform custom packet processing at a very early stage of the packet handling, limiting overheads and thus providing high-performance. At the same time, XDP allows programmers to leverage the Linux's kernel, e.g., selecting a subset of packets that should be processed by its network stack, which helps with compatibility and ease of development. XDP is part of the Linux kernel since release 4.18, and it is widely used in production environments~\cite{Cloudflare, Katran, tu2018bringing}. In most of these use cases, e.g., load balancing~\cite{Katran} and packet filtering~\cite{Cloudflare}, a majority of the received network packets is processed entirely within XDP. The production deployments of XDP have also pushed developers to optimize and minimize the XDP overheads, which now appear to be mainly related to the Linux driver model, as thoroughly discussed in ~\cite{xdp}. 

\begin{figure}[t]
	\begin{center}
	\includegraphics[width=1\columnwidth]{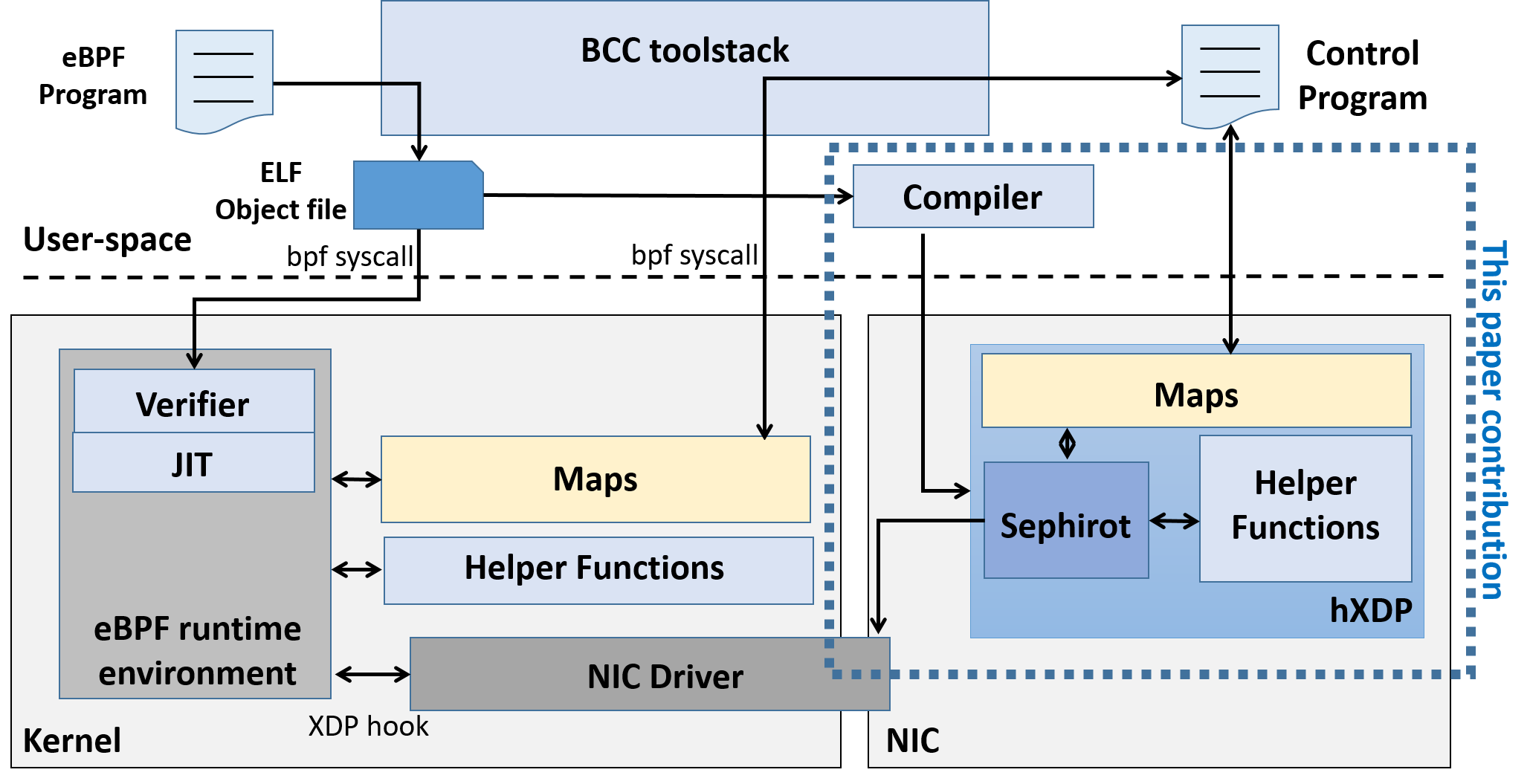} 
	\vspace{-18pt}
	\caption{An overview of the XDP workflow and architecture, including the contribution of this paper.} 
	\label{fig:overview}
    \end{center}
	\vspace{-26pt}
\end{figure}

XDP programs are based on the Linux's eBPF technology. eBPF provides an in-kernel virtual machine for the sandboxed execution of small programs within the kernel context. An overview of the eBPF architecture and workflow is provided in Figure~\ref{fig:overview}.
In its current version, the eBPF virtual machine has 11 64b registers: $r0$ holds the return value from in-kernel functions and programs, $r1-r5$ are used to store arguments that are passed to in-kernel functions, $r6-r9$ are registers that are preserved during function calls and $r10$ stores the frame pointer to access the stack. The eBPF virtual machine has a well-defined ISA composed of more than 100 fixed length instructions (64b). The instructions give access to different functional units, such as ALU32, ALU64, branch and memory.
Programmers usually write an eBPF program using the C language with some restrictions, which simplify the static verification of the program. Examples of restrictions include forbidden unbounded cycles, limited stack size, lack of dynamic memory allocation, etc. 

To overcome some of these limitations, eBPF programs can use \texttt{helper functions} that implement some common operations, such as checksum computations, and provide access to protected operations, e.g., reading certain kernel memory areas. 
eBPF programs can also access kernel memory areas called \texttt{maps}, i.e., kernel memory locations that essentially resemble tables. 
Maps are declared and configured at compile time to implement different data structures, specifying the type, size and an ID. For instance, eBPF programs can use maps to implement arrays and hash tables. An eBPF program can interact with map's locations by means of pointer deference, for un-structured data access, or by invoking specific helper functions for structured data access, e.g., a lookup on a map configured as a hash table.
Maps are especially important since they are the only mean to keep state across program executions, and to share information with other eBPF programs and with programs running in user space. In fact, a map can be accessed using its ID by any other running eBPF program and by the control application running in user space.
User space programs can load eBPF programs and read/write maps either using the \texttt{libbf} library or frontends such as the BCC toolstack.  

XDP programs are compiled using LLVM or GCC, and the generated ELF object file is loaded trough the \texttt{bpf} syscall, specifying the \texttt{XDP hook}. Before the actual loading of a program, the kernel verifier checks if it is safe, then the program is attached to the hook, at the network driver level. Whenever the network driver receives a packet, it triggers the execution of the registered programs, which starts from a clean context.

\subsection{Challenges}
To grasp an intuitive understanding of the design challenge involved in supporting XDP on FPGA, we now consider the example of an XDP program that implements a simple stateful firewall for checking the establishment of bi-directional TCP or UDP flows, and to drop flows initiated from an external location. 
We will use this function as a running example throughout the paper, since despite its simplicity, it is a realistic and widely deployed function. 

The simple firewall first performs a parsing of the Ethernet, IP and Transport protocol headers to extract the flow's 5-tuple (IP addresses, port numbers, protocol). Then, depending on the input port of the packet (i.e., external or internal) it either looks up an entry in a hashmap, or creates it. The hashmap key is created using an absolute ordering of the 5 tuple values, so that the two directions of the flow will map to the same hash. Finally, the function forwards the packet if the input port is internal or if the hashmap lookup retrieved an entry, otherwise the packet is dropped. A C program describing this simple firewall function is compiled to 71 eBPF instructions.

We can build a rough idea of the potential best-case speed of this function running on an FPGA-based eBPF executor, assuming that each eBPF instruction requires 1 clock cycle to be executed, that clock cycles are not spent for any other operation, and that the FPGA has a 156MHz clock rate, which is common in FPGA NICs~\cite{netsume}. In such a case, a naive FPGA implementation that implements the sequential eBPF executor would provide a maximum throughput of 2.8 Million packets per second (Mpps).\footnote{I.e., the FPGA can run 156M instructions per second, which divided by the 55 instructions of the program's expected execution path gives a 2.8M program executions per second. Here, notice that the execution path comprises less instructions that the overall program, since not all the program's instructions are executed at runtime due to if-statements.} Notice that this is a very optimistic upper-bound performance, which does not take into account other, often unavoidable, potential sources of overhead, such as memory access, queue management, etc. For comparison, when running on a single core of a high-end server CPU clocked at 3.7GHz, and including also operating system overhead and the PCIe transfer costs, the XDP simple firewall program achieves a throughput of 7.4 Million packets per second (Mpps).~\footnote{Intel Xeon E5-1630v3, Linux kernel v.5.6.4.} 
Since it is often undesired or not possible to increase the FPGA clock rate, e.g., due to power constraints, in the lack of other solutions the FPGA-based executor would be 2-3x slower than the CPU core.

Furthermore, existing solutions to speed-up sequential code execution, e.g., superscalar architectures, are too expensive in terms of hardware resources to be adopted in this case. In fact, in a superscalar architecture the speed-up is achieved leveraging instruction-level parallelism at runtime. However, the complexity of the hardware required to do so grows exponentially with the number of instructions being checked for parallel execution. This rules out re-using general purpose soft-core designs, such as those based on RISC-V~\cite{heinz2019catalog, pulpino}. 

\subsection{\nxdp Overview}
\nxdp addresses the outlined challenge by taking a software-hardware co-design approach. In particular, \nxdp provides both a compiler and the corresponding hardware module. The compiler takes advantage of eBPF ISA optimization opportunities, leveraging \nxdp's hardware module features that are introduced to simplify the exploitation of such opportunities. Effectively, we design a new ISA that extends the eBPF ISA, specifically targeting the execution of XDP programs.

The compiler optimizations perform transformations at the eBPF instruction level: remove unnecessary instructions; replace instructions with newly defined more concise instructions; and parallelize instructions execution. All the optimizations are performed at compile-time, moving most of the complexity to the software compiler, thereby reducing the target hardware complexity.
We describe the optimizations and the compiler in Section~\ref{sec:instruction-analysis}. 
Accordingly, the \nxdp hardware module implements an infrastructure to run up to 4 instructions in parallel, implementing a Very Long Instruction Word (VLIW) soft-processor. The VLIW soft-processor does not provide any runtime program optimization, e.g., branch prediction, instruction re-ordering, etc. We rely entirely on the compiler to optimize XDP programs for high-performance execution, thereby freeing the hardware module of complex mechanisms that would use more hardware resources. We describe the \nxdp hardware design in Section~\ref{sec:nxdp}.

Ultimately, the \nxdp hardware component is deployed as a self-contained IP core module to the FPGA. The module can be interfaced with other processing modules if needed, or just placed as a bump-in-the-wire between the NIC's port and its PCIe driver towards the host system. The \nxdp software toolchain, which includes the compiler, provides all the machinery to use \nxdp within a Linux operating system. 

From a programmer perspective, a compiled eBPF program could be therefore interchangeably executed in-kernel or on the FPGA, as shown in Figure~\ref{fig:overview}.\footnote{The choice of where to run an XDP program should be explicitly taken by the user, or by an automated control and orchestration system, if available.}

\section{\nxdp Compiler}
\label{sec:instruction-analysis}
In this section we describe the \nxdp instruction-level optimizations, and the compiler design to implement them.

\subsection{Instructions reduction}\label{instr_reduction}
The eBPF technology is designed to enable execution within the Linux kernel, for which it requires programs to include a number of extra instructions, which are then checked by the kernel's verifier. When targeting a dedicated eBPF executor implemented on FPGA, most such instructions could be safely removed, or they can be replaced by cheaper embedded hardware checks.
Two relevant examples are instructions for memory boundary checks and memory zero-ing.

\begin{figure}[t]
	\begin{center}
	\includegraphics[width=.9\columnwidth]{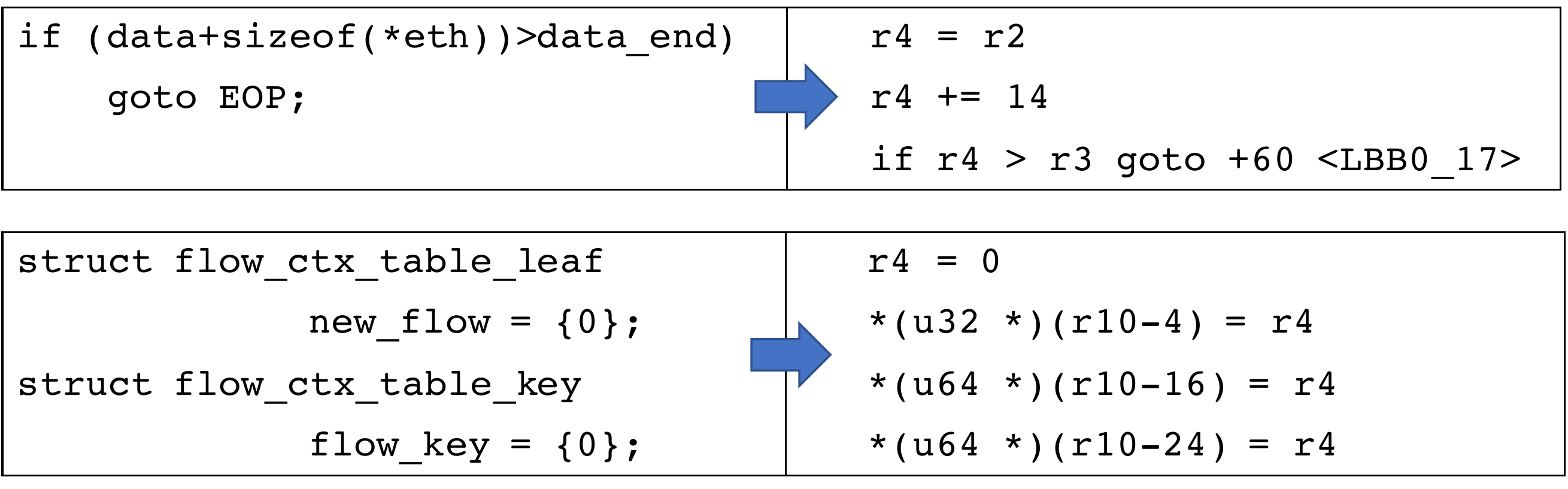} 
	\vspace{-6pt}
	\caption{Examples of instructions removed by hXDP} 
	\label{fig:listings-reduction}
    \end{center}
	\vspace{-12pt}
\end{figure}





\noindent\textbf{Boundary checks} are required by the eBPF verifier to ensure that programs only read valid memory locations, whenever a pointer operation is involved. For instance, this is relevant for accessing the socket buffer containing the packet data, during parsing.
Here, a required check is to verify that the packet is large enough to host the expected packet header. As shown in Figure~\ref{fig:listings-reduction}, a single check like this may cost 3 instructions, and it is likely that such checks are repeated multiple times. In the simple firewall case, for instance, there are three such checks for the Ethernet, IP and L4 headers.
In \nxdp we can safely remove these instructions, implementing the check directly in hardware.

\noindent\textbf{Zero-ing} is the process of setting a newly created variable to zero, and it is a common operation performed by programmers both for safety and for ensuring correct execution of their programs. A dedicated FPGA executor can provide hard guarantees that all relevant memory areas are zero-ed at program start, therefore making the explicit zero-ing of variables during initialization redundant. In the simple firewall function zero-ing requires 4 instructions, as shown in Figure  \ref{fig:listings-reduction}.

\subsection{ISA extension}\label{isa_ext}
To effectively reduce the number of instructions we define an ISA that enables a more concise description of the program. Here, there are two factors at play to our advantage. First, we can extend the ISA without accounting for constraints related to the need to support efficient Just-In-Time compilation. 
Second, our eBPF programs are part of XDP applications, and as such we can expect packet processing as the main program task. Leveraging these two facts we define a new ISA that changes in three main ways the original eBPF ISA.







\begin{figure}[t]
	\begin{center}
	\includegraphics[width=.9\columnwidth]{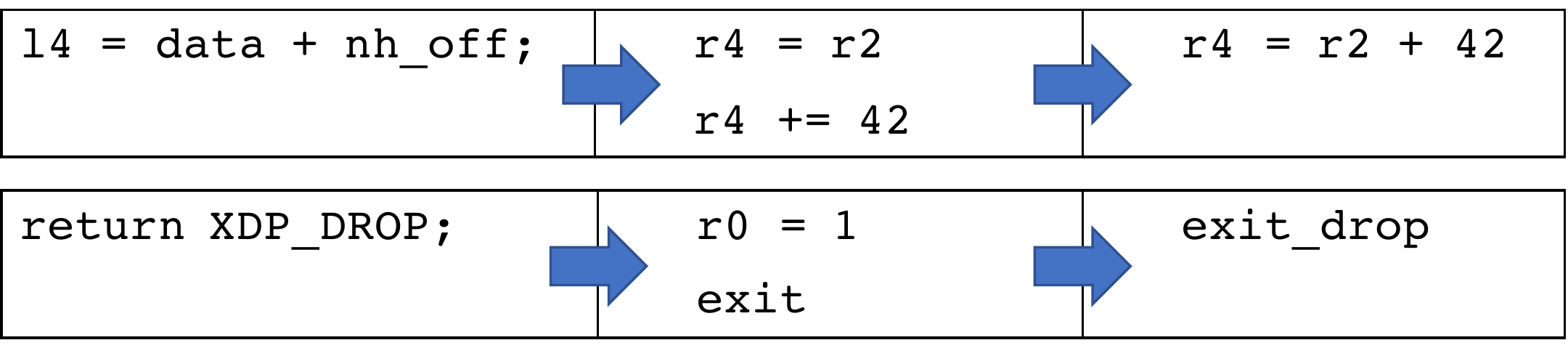} 
	\vspace{-6pt}
	\caption{Examples of hXDP ISA extensions} 
	\label{fig:listings-extension}
    \end{center}
	\vspace{-12pt}
\end{figure}

\noindent\textbf{Operands number}. The first significant change has to deal with the inclusion of three-operand operations, in place of eBPF's two-operand ones. Here, we believe that the eBPF's ISA selection of two-operand operations was mainly dictated by the assumption that an x86 ISA would be the final compilation target. Instead, using three-operand instructions allows us to implement an operation that would normally need two instructions with just a single instruction,  as shown in Figure~\ref{fig:listings-extension}. 

\noindent\textbf{Load/store size}. The eBPF ISA includes byte-aligned memory load/store operations, with sizes of 1B, 2B, 4B and 8B. While these instructions are effective for most cases, we noticed that during packet processing the use of 6B load/store can reduce the number of instructions in common cases. In fact, 6B is the size of an Ethernet MAC address, which is a commonly accessed field both to check the packet destination or to set a new one. 
Extending the eBPF ISA with 6B load/store instructions often halves the required instructions. 

\noindent\textbf{Parametrized exit}. The end of an eBPF program is marked by the exit instruction. In XDP, programs set the $r0$ to a value corresponding to the desired forwarding action (e.g., DROP, TX, etc), then, when a program exits the framework checks the $r0$ register to finally perform the forwarding action (see listing~\ref{fig:listings-extension}). While this extension of the ISA only saves one (runtime) instruction per program, as we will see in Section~\ref{sec:nxdp}, it will also enable more significant hardware optimizations.

\subsection{Instruction Parallelism}
Finally, we explore the opportunity to perform parallel processing of an eBPF program's instructions. Here, it is important to notice that high-end \textit{superscalar} CPUs are usually capable to execute multiple instructions in parallel, using a number of complex mechanisms such as speculative execution or out-of-order execution. However, on FPGAs the introduction of such mechanisms could incur significant hardware resources overheads. Therefore, we perform only a static analysis of the instruction-level parallelism of eBPF programs. 


To determine if two or more instructions can be parallelized, the three Bernstein conditions have to be checked~\cite{bern}. Simplifying the discussion to the case of two instructions $P_1, P_2$:
\begin{equation}
    I_1 \cap O_2 = \emptyset; O_1 \cap I_2 = \emptyset; O_2 \cap O_1 = \emptyset;
\end{equation}
Where $I_1, I_2$ are the instructions' input sets (e.g. source operands and memory locations) and $O_1, O_2$ are their output sets.
The first two conditions imply that if any of the two instructions depends on the results of the computation of the other, those two instructions cannot be executed in parallel. The last condition implies that if both instructions are storing the results on the same location, again they cannot be parallelized.
Verifying the Bernstein conditions and parallelizing instructions requires the design of a suitable compiler, which we describe next.

\subsection{Compiler design}
\label{sec:compiler}
We design a custom compiler to implement the optimizations outlined in this section and to transform XDP programs into a schedule of parallel instructions that can run with \nxdp.
The schedule can be visualized as a virtually infinite set of rows, each with multiple available spots, which need to be filled with instructions. The number of spots corresponds to the number of execution lanes of the target executor. The final objective of the compiler is to fit the given XDP program's instructions in the smallest number of rows.
To do so, the compiler performs five steps.

\noindent\textbf{\textit{Control Flow Graph} construction} First, the compiler performs a forward scan of the eBPF bytecode to identify the program's \textit{basic blocks}, i.e., sequences of instructions that are always executed together. 
The compiler identifies the first and last instructions of a block, and the control flow between blocks, by looking at branching instructions and jump destinations.
With this information it can finally build the \textit{Control Flow Graph} (CFG), which represents the basic blocks as nodes and the control flow as directed edges connecting them.

\noindent\textbf{\textit{Peephole} optimizations}
Second, for each basic block the compiler performs the removal of unnecessary instructions (cf. Section \ref{instr_reduction}), and the substitution of groups of eBPF instructions with an equivalent instruction of our extended ISA (cf. Section \ref{isa_ext}).


\noindent\textbf{\textit{Data Flow dependencies}} 
Third, the compiler discovers \textit{Data Flow dependencies}. This is done by implementing an iterative algorithm to analyze the CFG. The algorithm analyzes each block, building a data structure containing the block's input, output, defined, and used symbols. Here, a symbol is any distinct data value defined (and used) by the program. 
Once each block has its associated set of symbols, the compiler can use the CFG to compute data flow dependencies between instructions. This information is captured in per-instruction \textit{data dependency graphs} (DDG).

\noindent\textbf{Instruction scheduling}
Fourth, the compiler uses the CFG and the learned DDGs to define an instruction schedule that meets the first two Bernstein conditions. Here, the compiler takes as input the maximum number of parallel instructions the target hardware can execute, and potential hardware constraints it needs to account for. For example, as we will see in Section~\ref{sec:nxdp}, the \nxdp executor has 4 parallel execution lanes, but helper function calls cannot be parallelized. 

To build the instructions schedule, the compiler considers one basic block at a time, in their original order in the CFG. 
For each block, the compiler assigns the instructions to the current schedule's row, starting from the first instruction in the block and then searching for any other \textit{enabled} instruction. An instruction is enabled for a given row when its data dependencies are met, and when the potential hardware constraints are respected. E.g., an instruction that calls a helper function is not enabled for a row that contains another such instruction. If the compiler cannot find any enabled instruction for the current row, it creates a new row.
The algorithm continues until all the block's instructions are assigned to a row. 

At this point, the compiler uses the CFG to identify potential candidate blocks whose instructions may be added to the schedule being built for the current block. That is, such block's instructions may be used to fill gaps in the current schedule's rows.
The compiler considers as candidate blocks the current block's \textit{control equivalent blocks}. I.e, those blocks that are surely going to be executed if the current block is executed. Instructions from such blocks are checked and, if enabled, they are added to the currently existing schedule's rows. This allows the compiler to move in the current block's schedule also a series of branching instructions that are immediately following the current block, enabling a \textit{parallel branching} optimization in hardware (cf. Section~\ref{sec:nxdp-opt}).

When the current block's and its candidate blocks' enabled instructions are all assigned, the algorithm moves to the next block with instructions not yet scheduled, re-applying the above steps. The algorithm terminates once all the instructions have been assigned to the schedule. 

\noindent\textbf{Physical register assignment} 
Finally, in the last step the compiler assigns physical registers to the program's symbols. First, the compilers assigns registers that have a precise semantic, such as \texttt{r0} for the exit code, \texttt{r1}-\texttt{r5} for helper function argument passing, and \texttt{r10} for the frame pointer.
After these fixed assignment, the compiler checks if for every row also the third Bernstein condition is met, otherwise it renames the registers of one of the conflicting instructions, and propagates the renaming on the following dependant instructions.

\section{Hardware Module}
\label{sec:nxdp}

We design \nxdp as an independent IP core, which can be added to a larger FPGA design as needed. Our IP core comprises the elements to execute all the XDP functional blocks on the NIC, including \texttt{helper functions} and \texttt{maps}. This enables the execution of a program entirely on the FPGA NIC and therefore it avoids as much as possible PCIe transfers.

\begin{figure}[t]
	\begin{center}
	\includegraphics[width=1\columnwidth]{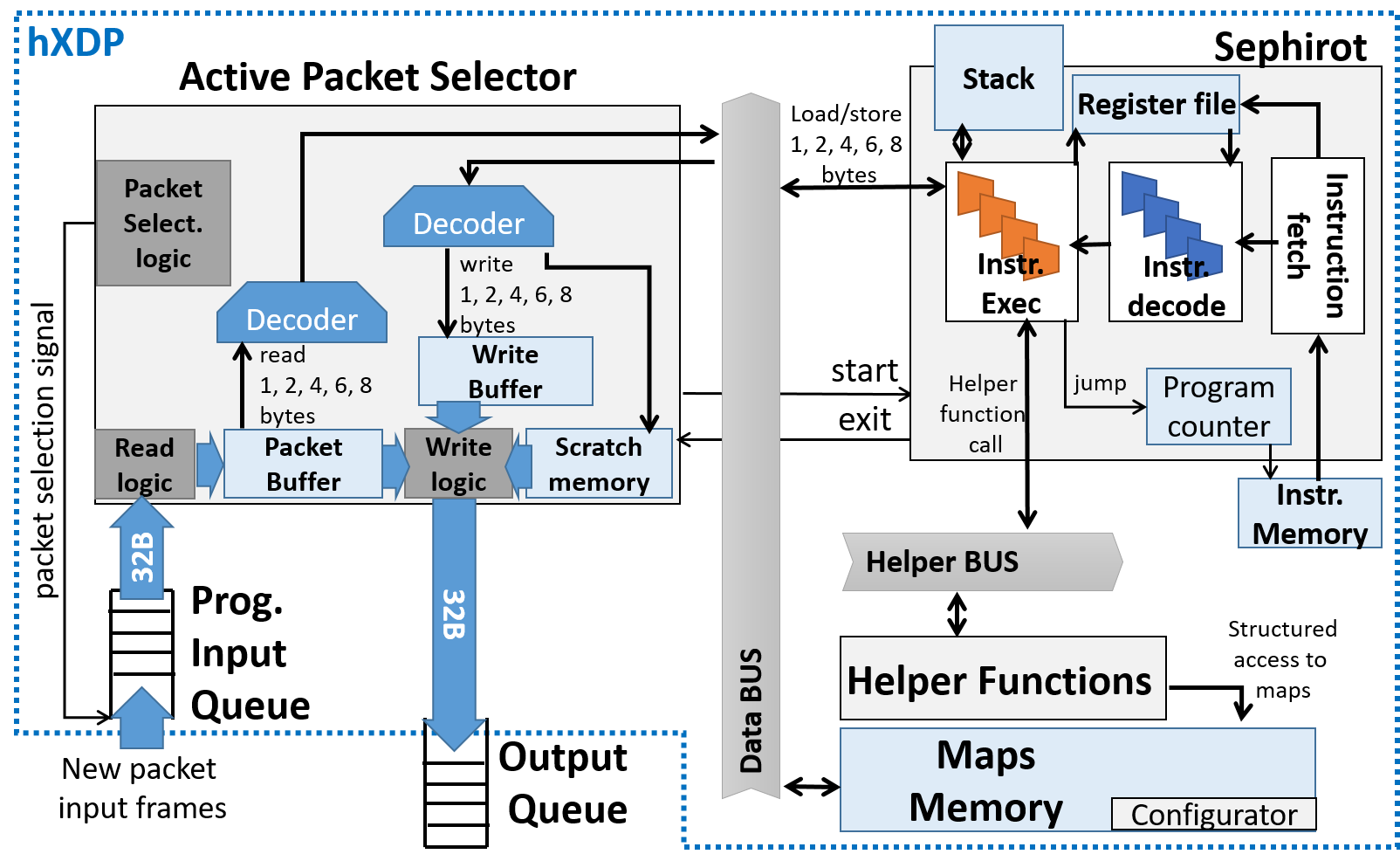} 
	\vspace{-12pt}
	\caption{The logic architecture of the \nxdp hardware design.} 
	\label{fig:hw}
    \end{center}
	\vspace{-20pt}
\end{figure}

\subsection{Architecture and components}
The \nxdp hardware design includes five components (see Figure~\ref{fig:hw}): the Programmable Input Queue (\texttt{PIQ}); the Active Packet Selector (\texttt{APS}); the \pproc processing core; the Helper Functions Module (\texttt{HF}); and the Memory Maps Module (\texttt{MM}). All the modules work in the same clock frequency domain. Incoming data is received by the PIQ. The APS reads a new packet from the PIQ into its internal packet buffer. In doing so, the APS provides a byte-aligned access to the packet data through a \textit{data bus}, which \pproc uses to selectively read/write the packet content. When the APS makes a packet available to the \pproc core, the execution of a loaded eBPF program starts. Instructions are entirely executed within \pproc, using 4 parallel execution lanes,  unless they call a helper function or read/write to maps. In such cases, the corresponding modules are accessed using the \textit{helper bus} and the \textit{data bus}, respectively. We detail each components next.
	
\subsubsection{Programmable Input queue}
When a packet is received, it enters the Programmable Input Queue (PIQ), which works as an interface with the NIC input bus. Thus, a packet is usually received divided into \textit{frames}, received at each clock cycle. The PIQ holds the packet's frames maintaining a \textit{head frame pointer}. The frames of a given packet can be therefore read from the queue independently from the reception order.

\subsubsection{Active Packet Selector}

The APS implements a finite-state machine to handle the transfer of a selected packet's frames from the PIQ to an APS' internal buffer.\footnote{The policy for selecting a given packet from the PIQ is by default FIFO, although this can be changed to implement more complex mechanisms.}
The internal buffer is large enough to hold a full-sized packet.

While the packet is stored divided in frames, the APS provides a byte-aligned read/write access to the data, as required by the eBPF ISA. I.e., the APS implements an eBPF program's \textit{packet buffer}, and builds the hardware-equivalent of the \texttt{xdp\_md} struct that is passed as argument to XDP programs. \pproc accesses such data structure using the main \nxdp's data bus.
Since \pproc has four parallel execution lanes, the APS provides four parallel read/write memory accesses through the data bus.

Storing the packet data in frames simplifies the buffer implementation. Nonetheless, this also makes the writing of specific bytes in the packet more complex. In particular, since only a frame-size number of bytes can be written to the buffer, the writing of single bytes would need to first read the entire frame, apply the single-byte modification, and then re-write to the buffer the entire modified frame. This is a complex operation, which would impact the maximum achievable clock rate if implemented in this way, or it would alternatively require multiple clock cycles to be completed. 
We instead use a \textit{difference buffer} to handle writes, trading off some memory space for hardware complexity. That is, modifications to the packet data are stored in a difference buffer that is byte addressed. As we will see next, the difference buffer allows us to separate the reading of certain packet data, which can be pre-fetched by \pproc during the decoding of an instruction, from the actual writing of new data in the packet, which can be performed during packet emission. 
In fact, the APS contains also a scratch memory to handle modifications to the packet that are applied before the current packet head. This is usually required by applications that use the \texttt{bpf\_adjust\_head} helper.

The scratch memory, the difference buffer, and the packet buffer are combined when \pproc performs a read of the packet data, and at the end of the program execution, during packet emission. Packet emission is the process of moving the modified packet data to the output queue. The entire process is handled by a dedicated finite-state machine, which is started by \pproc when an \textit{exit} instruction is executed. The emission of a packet happens in parallel with the reading of the next packet.

\subsubsection{\pproc}
\pproc is a VLIW processor with 4 parallel lanes that execute eBPF instructions. \pproc is designed as a pipeline of four stages: instruction fetch (IF); instruction decode (ID); instruction execute (IE); and commit. A program is stored in a dedicated instruction memory, from which \pproc fetches the instructions in order. The processor has another dedicated memory area to implement the program's stack, which is 512B in size, and 11 64b registers stored in the register file. These memory and register locations match one-to-one the eBPF virtual machine specification. \pproc begins execution when the APS has a new packet ready for processing, and it gives the processor \textit{start} signal.

On processor start (IF stage) a VLIW instruction is read and the 4 extended eBPF instructions that compose it are statically assigned to their respective execution lanes. In this stage, the  operands of the instructions are pre-fetched from the register file.
The remaining 3 pipeline stages are performed in parallel by the four execution lanes. 
During ID, memory locations are pre-fetched, if any of the eBPF instructions is a \textit{load}, while at the IE stage the relevant sub-unit are activated, using the relevant pre-fetched values. The sub-units are the Arithmetic and Logic Unit (ALU), the Memory Access Unit and the Control Unit. The ALU implements all the operations described by the eBPF ISA, with the notable difference that it is capable of performing operations on three operands. The memory access unit abstracts the access to the different memory areas, i.e., the stack, the packet data stored in the APS, and the maps memory. The control unit provides the logic to modify the program counter, e.g., to perform a \textit{jump}, and to invoke helper functions.
Finally, during the commit stage the results of the IE phase are stored back to the register file, or to one of the memory areas. \pproc terminates execution when it finds an exit instruction, in which case it signals to the APS the packet forwarding decision.

\subsubsection{Helper Functions}
\nxdp implements the XDP helper functions in a dedicated sub-module. We decided to provide a dedicated hardware implementation for the helper functions since their definition is rather static, and it changes seldom when new versions of the XDP technology are released. This also allows us to leverage at full the FPGA hardware parallelism to implement some more expensive functions, such as checksum computations. In terms of interface, the helper function sub-module offers the same interface provided by eBPF, i.e., helper functions arguments are read from registers r1-r5, and the return value is provided in r0. All values are exchanged using the dedicated helper data bus. Here, it is worth noticing that there is a single helper functions sub-module, and as such only one instruction per cycle can invoke a helper function.\footnote{Adding more sub-modules would not be sufficient to improve parallelism in this case, since we would need to also define additional registers to hold arguments/return values and include register renaming schemes. Adding sub-modules proved to be not helpful for most use cases.}
Among the helper functions there are the map lookup functions, which are used to implement hashmap and other data structures on top of the maps memory. Because of that, the helper functions sub-module has a direct access to the maps module.

\subsubsection{Maps}
The maps subsystem main function is to decode memory addresses, i.e., map id and row, to access the corresponding map's memory location. Here, one complication is that eBPF maps can be freely specified by a program, which defines the map's type and size for as many maps as needed. 
To replicate this feature in the hardware, the maps subsystem implements a \textit{configurator} which is instructed at program's load time. In fact, all the maps share the same FPGA memory area, which is then shaped by the configurator according to the maps section of the eBPF program, which (virtually) creates the appropriate number of maps with their row sizes, width and hash functions, e.g., for implementing hashmaps.

Since in eBPF single maps entries can be accessed directly using their address, the maps subsystem is connected via the data bus to \pproc, in addition to the direct connection to the helper function sub-module, which is instead used for structured map access. To enable parallel access to the \pproc's execution lanes, like in the case of the APS, the maps modules provides up to 4 read/write parallel accesses.

\subsection{Pipeline Optimizations}
\label{sec:nxdp-opt}

\vspace{0.1cm}
\noindent\textbf{Early processor start}
The packet content is transferred one frame per clock cycle from the PIQ to the APS.
Starting program execution without waiting the full transfer of the packet may trigger the reading of parts of it that are not yet transferred. 
However, handling such an exception requires only little additional logic to pause the \pproc pipeline, when the exception happens. In practice, XDP programs usually start reading the beginning of a packet, in fact in our tests we never experienced a case in which we had to pause \pproc. This provides significant benefits with packets of larger sizes, effectively masking the \pproc execution time.

\vspace{0.1cm}
\noindent\textbf{Program state self-reset}
As we have seen in Section~\ref{sec:instruction-analysis}, eBPF programs may perform zero-ing of the variables they are going to use. We provide automatic reset of the stack and of the registers at program initialization. This is an inexpensive feature in hardware, which improves security~\cite{costinSOSR20} and allows us to remove any such zero-ing instruction from the program.

\vspace{0.1cm}
\noindent\textbf{Data hazards reduction}
One of the issues of pipelined execution is that two instructions executed back-to-back may cause a race condition. If the first instruction produces a result needed by the second one, the value read by the second instruction will be stale, because of the operand/memory pretecthing performed by \pproc. 
Stalling the pipeline would avoid such race conditions at the cost of performance. Instead, we perform result forwarding on a per-lane basis. This allows the scheduling back-to-back of instructions on a single lane, even if the result of the first instruction is needed by the second one.\footnote{We empirically tested that a more complex intra-lane result forwarding does not provide measurable benefits.} The compiler is in charge of checking such cases and ensure that the instructions that have such dependancies are always scheduled on the same lane.

\begin{figure}[t]
	\begin{center}
	\includegraphics[width=.9\columnwidth]{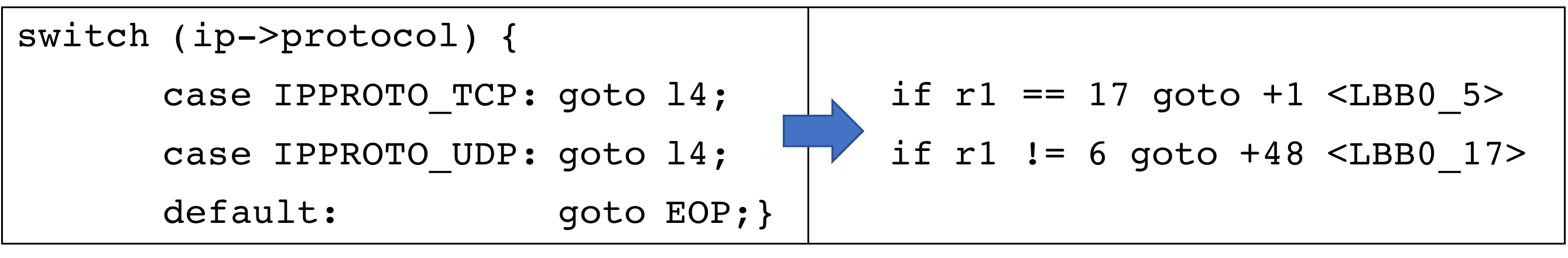} 
	\vspace{-6pt}
	\caption{Example of a switch statement} 
	\label{fig:listing-switch}
    \end{center}
	\vspace{-20pt}
\end{figure}

\vspace{0.1cm}
\noindent\textbf{Parallel branching}
The presence of branch instructions may cause performance problems with architectures that lack branch prediction, speculative and out of order execution. In the case of \pproc, this forces a serialization of the branch instructions.
However, in XDP programs there are often series of branches in close sequence, especially during header parsing (see Figure~\ref{fig:listing-switch}). 
We enabled the parallel execution of such branches, establishing a priority ordering of the \pproc's lanes. That is, all the branch instructions are executed in parallel by the VLIW's lanes. If more than one branch is taken, the highest priority one is selected to update the program counter. The compiler takes that into account when scheduling instructions, ordering the branch instructions accordingly.\footnote{This applies equally to a sequence of \texttt{if...else} or \texttt{goto} statements.}

%

\vspace{0.1cm}
\noindent\textbf{Early processor exit}
The processor stops when an exit instruction is executed. The exit instruction is recognized during the IF phase, which allows us to stop the processor pipeline early, and save the 3 remaining clock cycles. This optimization improves also the performance gain obtained by extending the ISA with parametrized exit instructions, as described in Section~\ref{sec:instruction-analysis}. In fact, XDP programs usually perform a move of a value to $r0$, to define the forwarding action, before calling an exit. Setting a value to a register always needs to traverse the entire \pproc pipeline. Instead, with a parametrized exit we remove the need to assign a value to $r0$, since the value is embedded in a newly defined exit instruction. 

\subsection{Implementation}
We prototyped \nxdp using the NetFPGA~\cite{netsume}, a board embedding 4 10Gb ports and a Xilinx Virtex7 FPGA. The \nxdp implementation uses a frame size of 32B and is clocked at 156.25MHz. Both settings come from the standard configuration of the NetFPGA reference NIC design.

\begin{table}[t]

        \begin{center}
          \begin{smaller}
            \begin{sc}
              \begin{tabular}{|c|c|c|c|}
                \hline
								\textbf{Component} & \textbf{ Logic} & \textbf{ Registers} & \textbf{BRAM } \\

                \hline
                PIQ &       215, 0.05\%  & 58, <0.01\% & 6.5, 0.44\%  \\
				APS &       9k, 2.09\%  & 10k, 1.24\% & 4, 0.27\%  \\ 
				\pproc &       27k, 6.35\%  & 4k, 0,51\% & -  \\ 
				Instr Mem &       -  & - & 7.7, 0.51\% \\
				Stack &       1k, 0.24\%  & 136, 0.02\% & 16, 1,09\% \\
				HF Subsystem &       339, 0.08\%  & 150, 0.02\% & -\\
				Maps Subsystem &       5.8k, 1.35\%  & 2.5k, 0.3\% & 16, 1.09\%  \\
				\hline
				total & 42k, 9.91\% & 18k, 2.09\% & 50, 3.40\% \\
				w/ reference NIC & 80k, 18.53\% & 63k, 7.3\% & 214, 14,63\% \\
				\hline
              \end{tabular}
            \end{sc}
          \end{smaller}
        \end{center}
        \vspace{-0.1in}
        \caption{NetFPGA resources usage breakdown, each row reports actual number and percentage of the FPGA total resources (\#, \% tot). \nxdp requires about 15\% of the FPGA resources in terms of Slice Logic and Registers.}
        \label{tab:resources}
        \vspace{-16pt}
\end{table}

The \nxdp FPGA IP core takes 9.91\% of the FPGA logic resources, 2.09\% of the register resources and 3.4\% of the FPGA's available BRAM. The considered BRAM memory does not account for the variable amount of memory required to implement maps. A per-component breakdown of the required resources is presented in Table~\ref{tab:resources}, where for reference we show also the resources needed to implement a map with 64 rows of 64B each.
As expected, the APS and \pproc are the components that need more logic resources, since they are the most complex ones.
Interestingly, even somewhat complex helper functions, e.g., a helper function to implement a hashmap lookup (\texttt{HF Map Access}), have just a minor contribution in terms of required logic, which confirms that including them in the hardware design comes at little cost while providing good performance benefits, as we will see in Section~\ref{sec:evaluation}.
When including the NetFPGA's reference NIC design, i.e., to build a fully functional FPGA-based NIC, the overall occupation of resources grows to 18.53\%, 7.3\% and 14.63\% for logic, registers and BRAM, respectively. This is a relatively low occupation level, which enables the use of the largest share of the FPGA for other accelerators.  
\section{Evaluation}
\label{sec:evaluation}
We use a selection of the Linux's XDP example applications and two real world applications to perform the \nxdp evaluation. The Linux examples are described in Table~\ref{tab:examples}. The real-world applications are the simple firewall we used as running example, and the Facebook's Katran server load balancer~\cite{Katran}.
Katran is a high performance software load balancer that translates virtual addresses to actual server addresses using a weighted scheduling policy, and providing per-flow consistency. Furthermore, Katran collects several flow metrics, and performs IPinIP packet encapsulation.

Using these applications, we perform an evaluation of the impact of the compiler optimizations on the programs' number of instructions, and the achieved level of parallelism. Then, we evaluate the performance of our NetFPGA implementation. In addition, we run a large set of micro-benchmarks to highlight features and limitations of \nxdp.
We use the microbenchmarks also to compare the \nxdp prototype performance with a Netronome NFP4000 SmartNIC. Although the two devices target different deployment scenarios, this can provide further insights on the effect of the \nxdp design choices. Unfortunately, the NFP4000 offers only limited eBPF support, which does not allow us to run a complete evaluation. We further include a comparison of \nxdp to other FPGA NIC programming solutions, before concluding the section with a brief dicussion of the evaluation results.

\begin{table}[]{\footnotesize
\begin{center}
\begin{tabular}{ll}
\hline
\textbf{Program}    & \textbf{Description} \\
\hline
xdp1	& parse pkt headers up to IP, and XDP\_DROP  \\
xdp2	& parse pkt headers up to IP, and XDP\_TX \\
xdp\_adjust\_tail & receive pkt, modify pkt into ICMP pkt and XDP\_TX\\      
router\_ipv4	& parse pkt headers up to IP,\\
                & look up in routing table and forward (redirect)\\
rxq\_info (drop)	& increment counter and XDP\_DROP  \\
rxq\_info (tx)	& increment counter and XDP\_TX  \\
tx\_ip\_tunnel    & parse pkt up to L4, encapsulate and XDP\_TX    \\
redirect\_map    & output pkt from a specified interface (redirect)   \\
\hline
\end{tabular}
\end{center}}
\vspace{-0.2in}
\caption{Tested Linux XDP example programs.}
\label{tab:examples}
\end{table}

\subsection{Compiler}
\begin{figure}[t]
\centering
\includegraphics[width=1\columnwidth]{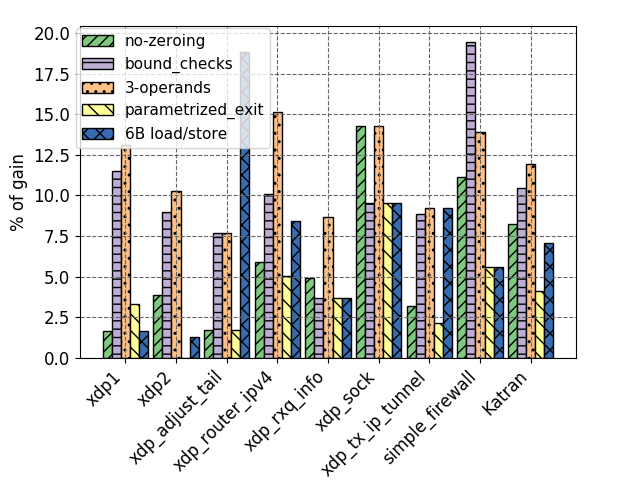}
	\caption{Reduction of instructions due to compiler optimizations, relative to the original number of instructions.} 
		\label{fig:optimizations}
		\vspace{-6pt}
\end{figure}

\noindent\textbf{Instruction-level optimizations} 
We evaluate the instruction-level optimizations described in Section~\ref{sec:instruction-analysis}, by activating selectively each of them in the \nxdp compiler.
Figure~\ref{fig:optimizations} shows the reduction of eBPF instructions for a program, relative to its original number of instructions. We can observe that the contribution of each optimization largely depends on the program. For instance, the \texttt{xdp\_adjust\_tail} performs several operations that need to read/write 6B of data, which in turn makes the 6B load/store instructions of our extended ISA particularly effective, providing a 18\% reduction in the number of instructions. Likewise, the \texttt{simple\_firewall} performs several bound checks, which account for 19\% of the program's instructions. 
The parametrized exit reduces the number of instructions by up to 5-10\%. However, it should be noted that this reduction has limited impact on the number of instructions executed at runtime, since only one exit instruction is actually executed.

\begin{figure}[t]
\centering
\includegraphics[width=1\columnwidth]{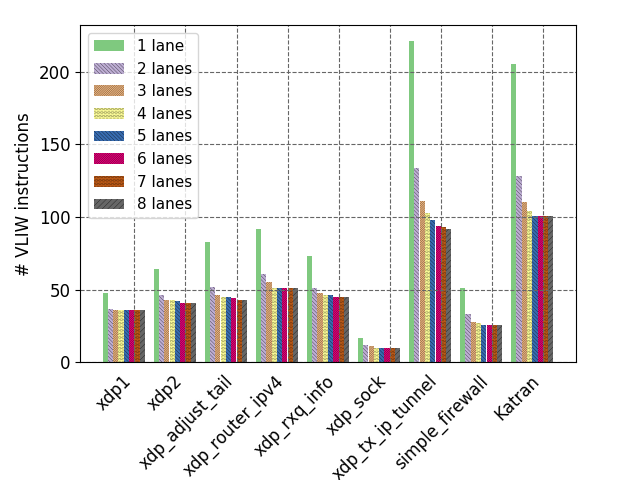} 
	\caption{Number of VLIW instructions when varying the available number of execution lanes.\\ } 
		\label{fig:parallelism}
		\vspace{-6pt}
\end{figure}

\begin{figure}[t]
\centering
\includegraphics[width=1\columnwidth]{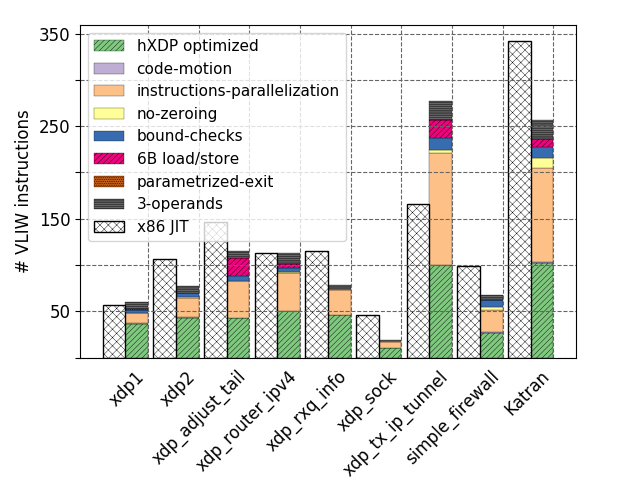} 
	\caption{Number of VLIW instructions, and impact of optimizations on its reduction.} 
		\label{fig:combined}
		\vspace{-6pt}
\end{figure}

\vspace{0.1cm}
\noindent\textbf{Instructions parallelization}
We configure the compiler to consider from 2 to 8 parallel execution lanes, and count the number of generated VLIW instructions. A VLIW instruction corresponds to a schedule's row (cf. Section~\ref{sec:compiler}), and it can therefore contain from 2 to 8 eBPF instructions in this test.
Figure~\ref{fig:parallelism} shows that the number of VLIW instructions is reduced significantly as we add parallel execution lanes up to 3, in all the cases. Adding a fourth execution lane reduces the VLIW instructions by an additional 5\%, and additional lanes provide only marginal benefits. Another important observation is that the compiler's physical register assignment step becomes more complex when growing the number of lanes, since there may not be enough registers to hold all the symbols being processed by a larger number of instructions.\footnote{This would ultimately require adding more registers, or the introduction of instructions to handle register spilling.} 
Given the relatively low gain when growing to more than four parallel lanes, we decided use four parallel lanes in \nxdp.

\vspace{0.1cm}
\noindent\textbf{Combined optimizations}
Figure~\ref{fig:combined} shows the final number of VLIW instructions produced by the compiler. We show the reduction provided by each optimization as a stacked column, and report also the number of x86 instructions, which result as output of the Linux's eBPF JIT compiler. In Figure, we report the gain for instruction parallelization, and the additional gain from \textit{code movement}, which is the gain obtained by anticipating instructions from control equivalent blocks (cf. Section~\ref{sec:compiler}). As we can see, when combined, the optimizations do not provide a simple sum of their gains, since each optimization affects also the instructions touched by the other optimizations. Overall, the compiler is capable of providing a number of VLIW instructions that is often 2-3x smaller than the original program's number of instructions. Notice that, by contrast, the output of the JIT compiler for x86 usually grows the number of instructions.\footnote{This is also due to the overhead of running on a shared executor, e.g., calling helper functions requires several instructions.}

\subsection{Hardware performance}
We compare \nxdp with XDP running on a server machine, and with the XDP offloading implementation provided by a SoC-based Netronome NFP 4000 SmartNIC. The NFP4000 has 60 programmable network processing cores (called \textit{microengines}), clocked at 800MHz. The server machine is 
equipped with an Intel Xeon E5-1630 v3 @3.70GHz, an Intel XL710 40GbE NIC, and running Linux v.5.6.4 with the i40e Intel NIC drivers. During the tests we use different CPU frequencies, i.e., 1.2GHz, 2.1GHz and 3.7GHz, to cover a larger spectrum of deployment scenarios. In fact, many deployments favor CPUs with lower frequencies and a higher number of cores~\cite{fbml}.
We use a DPDK packet generator to perform throughput and latency measurements. The packet generator is capable of generating a 40Gbps throughput with any packet size and it is connected back-to-back with the system-under-test, i.e., the \nxdp prototype running on the NetFPGA, the Netronome SmartNIC or the Linux server running XDP. Delay measurements are performed using hardware packet timestamping at the traffic generator's NIC, and measure the round-trip time. 
Unless differently stated, all the tests are performed using packets with size 64B belonging to a single network flow. This is a challenging workload for the systems under test.
Since we are interested in measuring the \nxdp hardware implementation performance, we do not perform tests that require moving packets to the host system. In such cases the packet processing performance would be largely affected by the PCIe and Linux drivers implementations, which are out-of-scope for this paper. We use a similar approach when running tests with the Netronome SmartNIC. As already mentioned, in this case we are only able to run a subset of the evaluation, i.e., some microbenchmarks, due to the the limited eBPF support implemented by the Netronome SmartNICs.


%


\begin{figure*}[!h]
	\centering
	\begin{tabularx}{\linewidth}{XXX}
	\includegraphics[width=0.60\columnwidth]{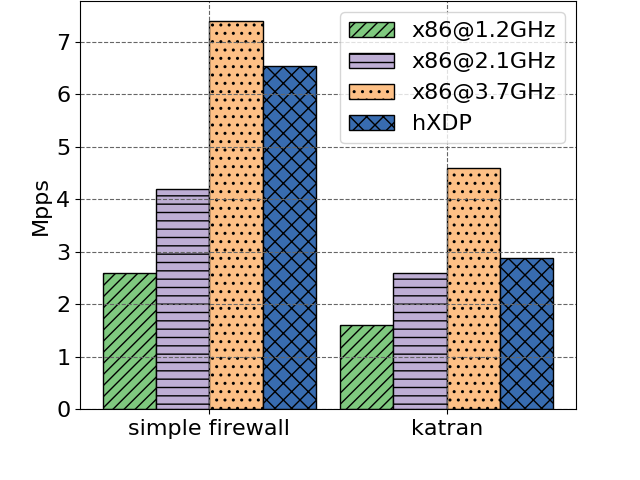} 
	\caption{Throughput for real-world applications. \nxdp is faster than a high-end CPU core clocked at over 2GHz.} 
		\label{fig:apptput}
		&
	\includegraphics[width=0.60\columnwidth]{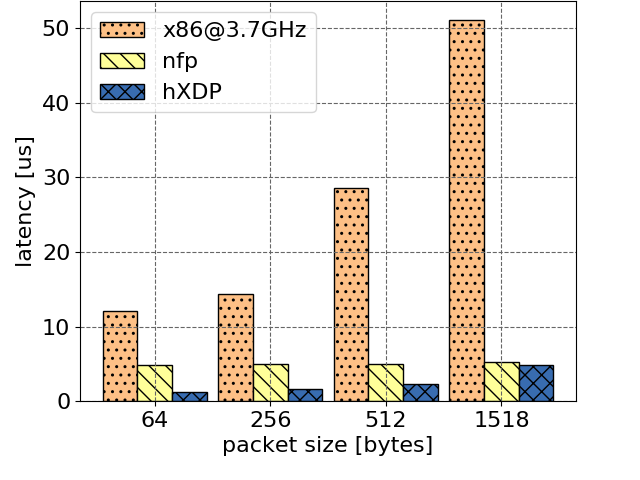} 
	\caption{Packet forwarding latency for different packet sizes. } 
		\label{fig:latency}
		&
	\includegraphics[width=0.60\columnwidth]{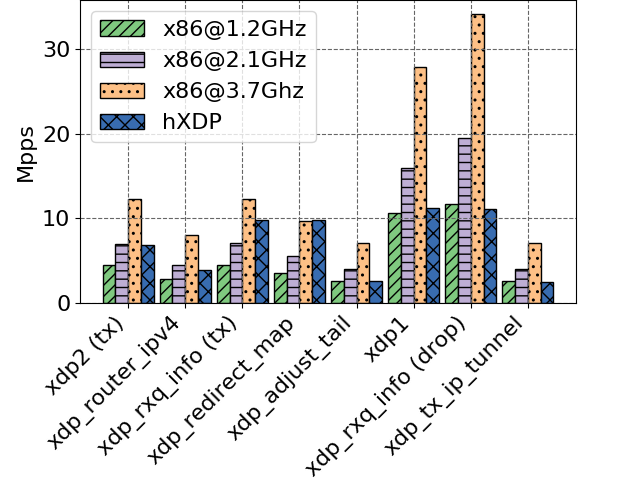} 
	\caption{Throughput of Linux's XDP programs. \nxdp is faster for programs that perform TX or redirection.} 
		\label{fig:ksamples}
	\end{tabularx}
	\vspace{-0.8cm}
\end{figure*}

\subsubsection{Applications performance}

\noindent\textbf{Simple firewall} 
In Section~\ref{sec:overview} we mentioned that an optimistic upper-bound for the hardware performance would have been 2.8Mpps. When using \nxdp with all the compiler and hardware optimizations described in this paper, the same application achieves a throughput of 6.53Mpps, as shown in Figure~\ref{fig:apptput}. This is only 12\% slower than the same application running on a powerful x86 CPU core clocked at 3.7GHz, and 55\% faster than the same CPU core clocked at 2.1GHz. In terms of latency, \nxdp provides about 10x lower packet processing latency, for all packet sizes (see    Figure~\ref{fig:latency}). This is the case since \nxdp avoids crossing the PCIe bus and has no software-related overheads. We omit latency results for the remaining applications, since they are not significantly different.\footnote{The impact of different programs is especially significant with small packet sizes. However, even in such cases we cannot observe significant differences. In fact each VLIW instruction takes about 7 nanoseconds to be executed, thus, differences of tens of instructions among programs change the processing latency by far less than a microsecond.} 
While we are unable to run the simple firewall application using the Netronome's eBPF implementation, Figure~\ref{fig:latency} shows also the forwarding latency of the Netronome NFP4000 (\texttt{nfp} label) when programmed with an XDP program that only performs packet forwarding. Even in this case, we can see that \nxdp provides a lower forwarding latency, especially for packets of smaller sizes. 

\vspace{0.1cm}
\noindent\textbf{Katran}
When measuring Katran we find that \nxdp is instead 38\% slower than the x86 core at 3.7GHz, and only 8\% faster than the same core clocked at 2.1GHz. The reason for this relatively worse \nxdp performance is the overall program length. Katran's program has many instructions, as such executors with a very high clock frequency are advantaged, since they can run more instructions per second.
However, notice the clock frequencies of the CPUs deployed at Facebook's datacenters~\cite{fbml} have frequencies close to 2.1GHz, favoring many-core deployments in place of high-frequency ones. \nxdp clocked at 156MHz is still capable of outperforming a CPU core clocked at that frequency.

\vspace{0.1cm}
\noindent\textbf{Linux examples}
We finally measure the performance of the Linux's XDP examples listed in Table~\ref{tab:examples}. These applications allow us to better understand the \nxdp performance with programs of different types (see Figure~\ref{fig:ksamples}). We can identify three categories of programs. First, programs that forward packets to the NIC interfaces are faster when running on \nxdp. These programs do not pass packets to the host system, and thus they can live entirely in the NIC. For such programs, \nxdp usually performs at least as good as a single x86 core clocked at 2.1GHz. In fact, processing XDP on the host system incurs the additional PCIe transfer overhead to send the packet back to the NIC.
Second, programs that always drop packets are usually faster on x86, unless the processor has a low frequency, such as 1.2GHz. Here, it should be noted that such programs are rather uncommon, e.g., programs used to gather network traffic statistics receiving packets from a network tap.
Finally, programs that are long, e.g., \texttt{tx\_ip\_tunnel} has 283 instructions, are faster on x86. Like we noticed in the case of Katran, with longer programs the \nxdp's implementation  low clock frequency can become problematic.

\subsubsection{Microbenchmarks}

\begin{figure*}[!h]
	\centering
	\begin{tabularx}{\linewidth}{XXX}
    \includegraphics[width=.60\columnwidth]{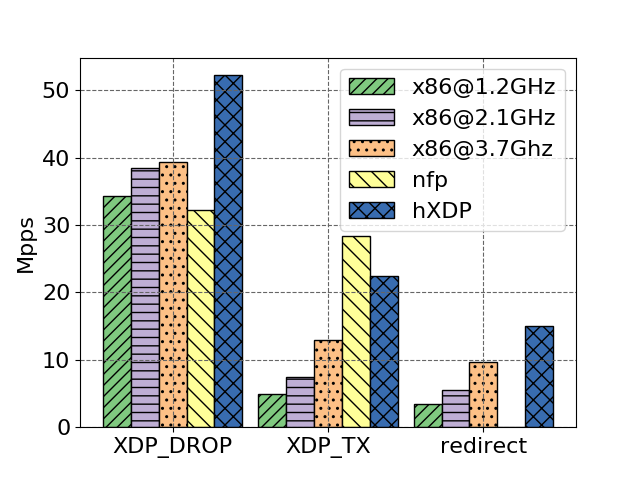} 
	\caption{Baseline throughput measurements for basic XDP programs.} 
	\label{fig:micro-tput}
		&
    \includegraphics[width=.60\columnwidth]{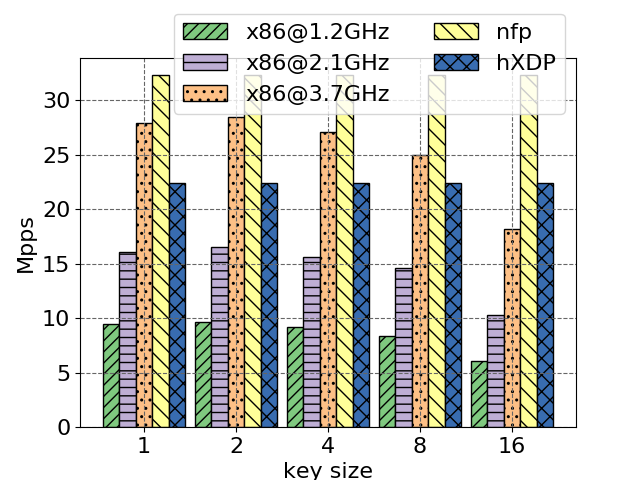}
	\caption{Impact on forwarding throughput of map accesses.} 
	\label{fig:memkey}
		&
    \includegraphics[width=.60\columnwidth]{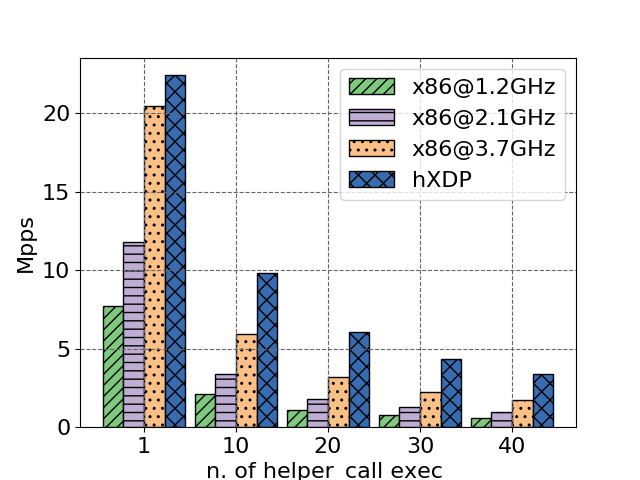} 
	\caption{Forwarding tput when calling a helper function.} 
		\label{fig:checksum}
	\end{tabularx}
	\vspace{-0.8cm}
\end{figure*}




\noindent\textbf{Baseline}
We measure the baseline packet processing performance using three simple programs: \textit{XDP\_DROP} drops the packet as soon as it is received; \textit{XDP\_TX} parses the Ethernet header and swaps MAC addresses before sending the packet out to the port from which it was received; \textit{redirect} is like XDP\_TX, but sends the packet out to a different port, which requires calling a specific XDP helper function.
The performance results clearly show the advantage of running on the NIC and avoiding PCIe transfers when processing small programs (see Figure~\ref{fig:micro-tput}). \nxdp can drop 52Mpps vs the 38Mpps of the x86 CPU core@3.7GHz, and 32Mpps of the Netronome NFP4000. Here, the very high performance of \nxdp is due to the parametrized exit/early exit optimizations mentioned in Section~\ref{sec:nxdp}. Disabling the optimization brings down the \nxdp performance to 22Mpps. In the case of XDP\_TX, instead, \nxdp forwards 22.5Mpps while x86 can forward 12Mpps. The NFP4000 is the fastest device in this test, forwarding over 28Mpps. In the case of redirect, \nxdp provides 15Mpps, while x86 can only forward 11Mpps when running at 3.7GHz. Here, the redirect has lower performance because eBPF implements it with a helper. Results for the NFP4000 are not available since the it does not support the redirect action.

\noindent\textbf{Maps access}
Accessing eBPF maps affects the performance of XDP programs that need to read and keep state. In this test we measure the performance variation when accessing a map with a variable key size ranging between 1-16B. Accessing the map is performed calling a helper function that performs a hash of the key and then retrieves the value from memory. In the x86 tests we ensure that the accessed entry is in the CPU cache. 
Figure~\ref{fig:memkey} shows that the \nxdp prototype has constant access performance, independently from the key size. This is  the result of the wider memory data buses, which can in fact accomodate a memory access in a single clock cycle for keys of up to 32B in size. The NFP4000, like \nxdp, shows no performance drop when increasing the key size. Instead, in the x86 case the performance drops when the key size grows from 8B to 16B, we believe this is due to the need to perform multiple loads.

\noindent\textbf{Helper functions}
In this micro-benchmark we measure throughput performance when calling from 1 to 40 times a helper function that performs an incremental checksum calculation. Since helper functions are implemented as dedicated hardware functions in \nxdp, we expect our prototype to exhibit better performance than x86, which is confirmed by our results (see Figure~\ref{fig:checksum}).
I.e., \nxdp may provide significant benefits when offloading programs that need complex computations captured in helper functions. Also, the  \nxdp's helper function machinery may be used to eventually replace sets of common instructions with more efficient dedicated hardware implementations, providing an easier pathway for future extensions of the hardware implementation.

\noindent\textbf{Instructions per cycle}
We compare the parallelization level obtained at compile time by \nxdp, with the runtime parallelization performed by the x86 CPU core. Table~\ref{tab:examples_ipc} shows that the static \nxdp parallelization achieves often a parallelization level as good as the one achieved by the complex x86 runtime machinery.\footnote{The x86 IPC should be understood as a coarse-grained estimation of the XDP instruction-level parallelism since, despite being isolated, the CPU runs also the operating system services related to the eBPF virtual machine, and its IPC is also affected by memory access latencies, which more significantly impact the IPC for high clock frequencies.}

\begin{table}[]{\footnotesize
\begin{center}
\begin{tabular}{llll}
\hline
\textbf{Program}    &  \textbf{\# instr.} &  \textbf{x86 IPC} & \textbf{\nxdp IPC} \\
\hline
xdp1                & 61  & 2.20 & 1.70 \\
xdp2	            & 78  & 2.19 & 1.81 \\
xdp\_adjust\_tail   & 117  &  2.37  & 2.72 \\
router\_ipv4        & 119 & 2.38 & 2.38 \\
rxq\_info           & 81  & 2.81 & 1.76 \\
tx\_ip\_tunnel      & 283  & 2.24 & 2.83 \\
simple\_firewall    & 72  & 2.16 &  2.66\\
Katran              & 268  & 2.32 & 2.62 \\

\hline
\end{tabular}
\end{center}}
\vspace{-12pt}
\caption{Programs' number of instructions, x86 runtime instruction-per-cycle (IPC) and \nxdp static IPC mean rates.}
\label{tab:examples_ipc}
\vspace{-0.2in}
\end{table}

\subsubsection{Comparison to other FPGA solutions}
\nxdp provides a more flexible programming model than previous work for FPGA NIC programming. However, in some cases, simpler network functions implemented with \nxdp could be also implemented using other programming approaches for FPGA NICs, while keeping functional equivalence. One such example is the simple firewall presented in this paper, which is supported also by FlowBlaze~\cite{FlowBlaze}. 

\vspace{0.1cm}
\noindent\textbf{Throughput} 
Leaving aside the cost of re-implementing the function using the FlowBlaze abstraction, we can generally expect \nxdp to be slower than FlowBlaze at processing packets. In fact, in the simple firewall case, FlowBlaze can forward about 60Mpps, vs the 6.5Mpps of \nxdp.
The FlowBlaze design is clocked at 156MHz, like \nxdp, and its better performance is due to the high-level of specialization. FlowBlaze is optimized to process only packet headers, using statically-defined functions. This requires loading a new bitstream on the FPGA when the function changes, but it enables the system to achieve the reported high performance.\footnote{FlowBlaze allows the programmer to perform some runtime reconfiguration of the functions, however this a limited feature. For instance, packet parsers are statically defined.}
Conversely, \nxdp has to pay a significant cost to provide full XDP compatibility, including dynamic network function programmability and processing of both packet headers and payloads.

\vspace{0.1cm}
\noindent\textbf{Hardware resources} 
A second important difference is the amount of hardware resources required by the two approaches. \nxdp needs about 18\% of the NetFPGA logic resources, independently from the particular network function being implemented. Conversely, FlowBlaze implements a packet processing pipeline, with each pipeline's stage requiring about 16\% of the NetFPGA's logic resources. For example, the simple firewall function implementation requires two FlowBlaze pipeline's stages. More complex functions, such as a load balancer, may require 4 or 5 stages, depending on the implemented load balancing logic~\cite{FlowBlaze-repo}.

\vspace{0.1cm}
In summary, the FlowBlaze's pipeline leverages hardware parallelism to achieve high performance. However, it has the disadvantage of often requiring more hardware resources than a sequential executor, like the one implemented by \nxdp. Because of that, \nxdp is especially helpful in scenarios where a small amount of FPGA resources is available, e.g., when sharing the FPGA among different accelerators.

\subsection{Discussion}
\noindent\textbf{Suitable applications} \nxdp can run XDP programs with no modifications, however, the results presented in this section show that \nxdp is especially suitable for programs that can process packets entirely on the NIC, and which are no more than a few 10s of VLIW instructions long. This is a common observation made also for other offloading solutions~\cite{hohlfeld2019demystifying}.

\vspace{0.1cm}
\noindent\textbf{FPGA Sharing} 
At the same time, \nxdp succeeds in using little FPGA resources, leaving space for other accelerators. For instance, we could co-locate on the same FPGA several instances of the VLDA accelerator design for neural networks presented in \cite{tvm}. 
Here, one important note is about the use of memory resources (BRAM). Some XDP programs may need larger map memories. It should be clear that the memory area dedicated to maps reduces the memory resources available to other accelerators on the FPGA. As such, the memory requirements of XDP programs, which are anyway known at compile time, is another important factor to consider when taking program offloading decisions.
    
\section{Future work}
While the \nxdp performance results are already good to run real-world applications, e.g., Katran, we identified a number of optimization options, as well as avenues for future research.

\vspace{0.1cm}
\noindent\textbf{Compiler} 
First, our compiler can be improved. For instance, we were able to hand-optimize the simple firewall instructions and run it at 7.1Mpps on \nxdp. This is almost a 10\% improvement over the result presented in Section~\ref{sec:evaluation}. The applied optimizations had to do with a better organization of the memory accesses, and we believe they could be automated by a smarter compiler. 

\vspace{0.1cm}
\noindent\textbf{Hardware parser} 
Second, XDP programs often have large sections dedicated to packet parsing. Identifying them and providing a dedicated programmable parser in hardware~\cite{programmableParser} may significantly reduce the number of instructions executed by \nxdp. However, it is still unclear what would be the best strategy to implement the parser on FPGA and integrate it with \nxdp, and the related performance and hardware resources usage trade offs.  

\vspace{0.1cm}
\noindent\textbf{Multi-core and memory}
Third, while in this paper we focused on a single processing core, \nxdp can be extended to support two or more \pproc cores. This would effectively trade off more FPGA resources for higher forwarding performance. For instance, we were able to test an implementation with two cores, and two lanes each, with little effort. This was the case since the two cores shared a common memory area and therefore there were no significant data consistency issues to handle. Extending to more cores (lanes) would instead require the design of a more complex memory access system. Related to this, another interesting extension to our current design would be the support for larger DRAM or HBM memories, to store large memory maps.

\vspace{0.1cm}
\noindent\textbf{ASIC}
Finally, \nxdp targets FPGA platforms, since we assume that FPGAs are already available in current deployments. Nonetheless, \nxdp has several fixed design's components, such as the \pproc core, which suggests that \nxdp may be implemented as ASIC. An ASIC could provide a potentially higher clock frequency, and an overall more efficient use of the available silicon resources. Here, in addition to measuring the performance that such a design may achieve, there are additional interesting open questions. For instance evaluating the potential advantages/disadvantages provided by the ability to change helper functions implemented in the FPGA, when compared to a fixed set of helper functions provided in ASIC.
\section{Related Work}
\label{sec:related}
\noindent\textbf{NIC Programming} 
AccelNet~\cite{AccelNet} is a match-action offloading engine used in large cloud datacenters to offload virtual switching and firewalling functions, implemented on top of the Catapult FPGA NIC~\cite{7783710}. 
FlexNIC \cite{flexNIC} is a design based on the RMT \cite{rmt} architecture, which  provides a flexible network DMA interface used by operating systems and applications to offload stateless packet parsing and classification.  P4->NetFPGA~\cite{p4netfpga} and P4FPGA~\cite{p4fpga} provide high-level synthesis from the P4~\cite{p4} domain-specific language to an FPGA NIC platform. FlowBlaze~\cite{FlowBlaze} implements a finite-state machine abstraction using match-action tables on an FPGA NIC, to implement simple but high-performance network functions.
Emu \cite{sultana2017emu} uses high level synthesis to implement functions described in C\# on the NetFPGA.
Compared to these works, instead of match-action or higher-level abstractions, \nxdp leverages abstractions defined by the Linux's kernel, and implements network functions described using the eBPF ISA. 

The Netronome SmartNICs implement a limited form of eBPF/XDP offloading~\cite{kicinski2016ebpf}. Unlike \nxdp that implements a solution specifically targeted to XDP programs, the Netronome solution is added on top of their network processor as an afterthought, and therefore it is not specialized for the execution of XDP programs. 

\vspace{0.1cm}
\noindent\textbf{Application frameworks} AccelTCP \cite{acceltcp}, Tonic \cite{arashloo2020enabling} and Xtra  \cite{bianchi2019xtra} present abstractions, hardware architectures and prototypes to offload the transport protocol tasks to the NIC. We have not investigated the feasibility of using \nxdp for a similar task, which is part of our future work. NICA~\cite{nica} and ClickNP~\cite{clickNP} are software/hardware frameworks that introduce specific software abstractions that connect FPGA blocks with an user program running on a general purpose CPU. In both cases, applications can only be designed composing the provided hardware blocks.
\nxdp provides instead an ISA that can be flexibly programmed, e.g., using higher level languages such as C.

\vspace{0.1cm}
\noindent\textbf{Applications} Examples of applications implemented on NICs include: DNS resolver~\cite{woodruff2019p4dns}; the paxos protocol~\cite{tokusashi2019case}; network slicing~\cite{yan2020p4}; Key-value stores~ \cite{li2017kv,lavasani2013fpga,tokusashi2018lake,siracusano2017smartnic}; Machine Learning~\cite{8416814, brainwave, 7477459}; and generic cloud services as proposed in \cite{sapio2017network,7783710, 10.5555/2665671.2665678}. 
\cite{liu2019e3} uses SmartNICs to provide a microservice-based platform to run different services. In this case, SoC-based NICs are used, e.g., based on Arm processing cores.
Lynx~\cite{lynx} provides a system to implement network-facing services that need access to accelerators, such as GPUs, without involving the host system's CPU.
Floem~\cite{phothilimthana2018floem} is a framework to simplify the design of applications that leverage offloading to SoC-based SmartNICs. \nxdp provides an XDP programming model that can be used to implement and extend these applications.

\vspace{0.1cm}
\noindent\textbf{NIC Hardware} 
Previous work presenting VLIW core designs for FPGAs did not focus on network processing~\cite{iseli1993spyder,jones2005fpga}. \cite{brunella2018v} is the closest to \nxdp. It employs a non-specialized MIPS-based ISA and a VLIW architecture for packet processing. \nxdp has an ISA design specifically targeted to network processing using the XDP abstractions. 
\cite{corundum} presents an open-source 100-Gbps FPGA NIC design. \nxdp can be integrated in such design to implement an open source FPGA NIC with XDP offloading support.
\section{Conclusion}
This paper presented the design and implementation of \nxdp, a system to run Linux's XDP programs on FPGA NICs. \nxdp can run unmodified XDP programs on FPGA matching the performance of a high-end x86 CPU core clocked at more than 2GHz.
Designing and implementing \nxdp required a significant research and engineering effort, which involved the design of a processor and its compiler, and while we believe that the performance results for a design running at 156MHz are already remarkable, we also identified several areas for future improvements. In fact, we consider \nxdp a starting point and a tool to design future interfaces between operating systems/applications and network interface cards/accelerators. To foster work in this direction, we make our implementations available to the research community.\footnote{\url{https://github.com/axbryd}}

\section*{Acknowledgments}
We would like to thank the anonymous reviewers and
our shepherd Costin Raiciu for their extensive and valuable feedback and comments, which have substantially improved
the content and presentation of this paper.

The research leading to these results has received funding from the ECSEL Joint Undertaking in collaboration with the European Union's H2020 Framework Programme (H2020/2014-2020) and National Authorities, under grant agreement n. 876967 (Project "BRAINE").

\bibliographystyle{abbrv}
\bibliography{biblio}

\section*{Appendix: Artifacts}

Source code, examples and instructions to replicate the results presented in this paper are provided at \url{https://github.com/axbryd/hXDP-Artifacts}.

\clearpage

\end{document}